\documentstyle[psfig]{aa}                     

\def\magcir{\raise -2.truept\hbox{\rlap{\hbox{$\sim$}}\raise5.truept
\hbox{$>$}\ }}

  \title{INTEGRAL observations of recurrent fast X-ray transient sources}
  \subtitle{}

   \author{ V. Sguera \inst{1}, E. J. Barlow\inst{1}, A. J. Bird\inst{1}, D. J. Clark\inst{1}, A. J. Dean\inst{1}, A. B. Hill\inst{1}, 
            L. Moran\inst{1}, S. E. Shaw\inst{1,2}, D. R. Willis\inst{1}, 
             A. Bazzano\inst{3}, P. Ubertini\inst{3},
            A. Malizia\inst{4}}

   \offprints{sguera@astro.soton.ac.uk}

   \institute{ School of Physics and Astronomy, University of Southampton, Highfield, SO17 1BJ, UK \and 
   INTEGRAL Science Data Center, CH-1290 Versoix, Switzerland \and  
   IASF/CNR, via Fosso del Cavaliere 100, 00133 Roma, Italy \and 
   IASF/CNR, via Piero Gobetti 101, I-40129 Bologna, Italy}

   \date{Received 21 March 2005 / Accepted 17 August 2005}

  \titlerunning{}

  \authorrunning{Sguera et al.}

\begin{document}

\abstract{Fast X-ray Transients (FXTs) are believed to be non-recurrent bright X-ray sources lasting  
less than a day and occuring  at serendipitous positions, they can best be
detected and discovered by instruments having a sufficiently wide field of view and high sensitivity. The IBIS/ISGRI instrument onboard INTEGRAL
is particularly suited to detect new or already known fast X-ray transient sources. We report on IBIS/ISGRI detection of newly
discovered outbursts of three fast transient sources located at low Galactic latitude: SAX~J1818.6$-$1703; IGR~J16479$-$4514; IGR~J17391$-$302/XTE~J1739$-$302.
The reported results confirm and strengthen the very fast transient nature of these sources, given that all their newly detected outbursts 
have a duration less than $\sim$ 3 hours. Additionally, they provide the first evidence for a 
possible recurrent fast transient behaviour as all three  
sources were detected in outburst by ISGRI more than once during the last 2 years.         
  

}
 \maketitle


\section{Introduction} 
The X-ray and $\gamma$-ray sky is characterized by marked variability
with many sources suddenly appearing and then disappearing again.
Amongst the many kinds of variable sources, X-ray transients have been observed 
since the first X-ray observations. According to their duration, they can be mainly divided into two
different classes: long duration 
and short duration X-ray transients. The former can be visible for weeks or even months and  are mainly related 
to Be-NS binaries systems (Be X-ray transients) or soft X-ray transients (a sub class of low mass X-ray binaries).
The latter last from hours to a few days and include, among  others,  a subclass of sources called Fast X-ray Transients (FXTs) which
last from a very few hours to less than a day.  
Recently, the term X-ray Flash (XRF) has been coined for a new subclass 
of FXTs lasting less than 1000 seconds (Heise et al. 2001), these are thought to be related to gamma ray bursts since  the two classes
share several observational properties.\\
Although FXTs  are among the most extreme examples of variability in the X-ray sky, 
not too much is known about them, mainly because of the lack of counterparts at other wavelengths.
They occur at unpredictable locations and times and it is very difficult to detect and observe them using traditional narrow field X-ray 
instruments. On the contrary, detectors having a sufficiently wide field of view are particularly suited to detect fast X-ray transient sources.
The larger the field of view of an instrument, the greater the chance of serendipitously detecting
a short duration random event, such as a FXT.\\
To date, only a few hundred FXTs have been observed by such wide field X-ray instruments, from HEAO-1 to BeppoSAX, 
although their all-sky rate is believed to be of the order of thousands per year (Arefiev V. A. et al. 2003).
Normally, FXTs are non-recurrent bright X-ray sources with no detectable quiescence emission;
this implies high peak-to-quiescence flux ratios (10$^{2}$--10$^{3}$)(Ambruster et al. 1983).
Less than half of all reported FXTs have been optically identified with stellar flares originating in active galactic coronal sources
as RS CVn binaries and dMe-dKe flare stars (Heise et al. 1975; Garcia et al. 1980; McHardy et al. 1982), Algol-type binaries 
(Schnopper et al. 1976, Favata 1998), W UMa systems and young T Tauri stars.
However, FXTs are a heterogeneous collection of objects showing large diversity of observational characteristics (light curves and spectra) 
which seem to suggest that these events are caused 
by more than one physical mechanism. In fact many unidentified FXTs do not appear to be coronal events, other progenitors have been 
suggested  such as: dwarf novae (Stern et al. 1981), ordinary and anomalous Type I X-ray bursts (Hoffman et al. 1978, Lewis $\&$ Joss 1981), 
BL Lac objects (Catanese et al. 1997; Maraschi et al. 1999).\\
The IBIS/ISGRI detector (Ubertini et al. 2003, Lebrun et al. 2003) on board the INTEGRAL satellite (Winkler et al. 2003) is 
particularly suited to detect new or already known fast X-ray transient sources.
Firstly, INTEGRAL provides a sensitivity higher than any earlier monitoring instruments and 
high resolution images with a large field of view fully coded (9$^{\circ}$$\times$9$^{\circ}$) 
as well as partially coded (full width at zero response, 30$^{\circ}$$\times$30$^{\circ}$). 
Furthermore, the INTEGRAL mission dedicates most of its
Core Program to deep exposures of the central Galactic region  and scans of the Galactic plane. 
To date, many new sources have been detected with IBIS/ISGRI, especially 
toward the inner region of the Galaxy.
The emerging picture is that they are High Mass X-ray Binaries (HMXBs), mainly Be/X-ray binaries (Lutovinov et al. 2004).
However, there are reasons to think 
that not all the new ISGRI sources are HMXB, especially those showing a very quick transient 
nature with activity period less than a day.

Previous to this work no recurring FXTs had been identified, 
in this paper  we report on INTEGRAL observations of three fast X-ray transient sources 
(SAX~J1818.6$-$1703, IGR~J16479$-$4514 and IGR~J17391$-$302/XTE~J1739$-$302)  located at low Galactic latitude and detected by ISGRI 
more than once during the last 2 years. They are 
characterized by extremely short outbursts (duration of few hours). We present results on several newly detected outbursts, not previously reported in the literature.
\section{INTEGRAL data analysis}
The reduction and analysis of the IBIS/ISGRI data have been performed by using the INTEGRAL Offline Scientific Analysis (OSA) v.4.1
available to the public through the INTEGRAL Science Data Centre ISDC (Courvoisier et al. 2003). 
INTEGRAL observations are typically divided into short pointings (Science Window, ScW) of  $\sim$ 2000 s duration.
 An analysis at the ScW level of the deconvolved ISGRI shadowgrams has been performed to search for transient 
sources detected (with a significance greater than  6$\sigma$) in a few consecutive ScWs  in the energy band 20--30~keV.
The ScW data set consists of all Core Program observations (the Galactic Plane Survey and
the Galactic Centre Deep Exposure).
This has been followed by  a timing analysis of the newly discovered outbursts, using OSA 4.1 to extract light curves and spectra.
Since coded mask instruments are characterized by large field of view (FOV), the brightest sources could affect the detection of the other nearby sources, especially of
the weaker ones. Bearing this in mind, we have also investigated the variability pattern of very bright sources in the FOV, 
besides the sources of  interest. They have shown a different time variability enabling us to conclude that the light curves  obtained 
for  the newly discovered outbursts  are reliable and uncontaminated.\\
The X-Ray Monitor JEM-X (Lund et al. 2003) on board the INTEGRAL satellite makes observations simultaneously with IBIS/ISGRI, providing images in the energy band 3--35~keV
with a 13$^\circ$.2 diameter field of view (although noise towards the edge of the FOV limits the usable area for weak sources to the central 
10$^\circ$.5).
Images from JEM-X were created for all newly discovered outbursts reported in this paper. 
In most cases, the source was outside or on the edge of the JEM-X field of view. 
With one exception (see section 5.1), no significant detections were obtained for  those inside the FOV.        
\section{SAX~J1818.6$-$1703}
\subsection{Archival X-ray observations of the source}
\begin{figure}[t!]
\psfig{figure=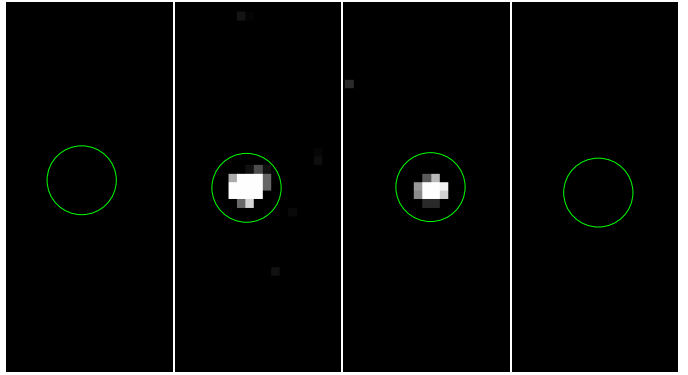,height=5cm,width=9cm}
\caption{ISGRI Science Window image sequence (20--30 keV) of a newly discovered  outburst (No. 1 in Table 1) of SAX~J1818.6$-$1703 (encircled). 
The duration of each ScW is $\sim$ 2000 s. 
Significance of the detection, left to right, was $<$2$\sigma$,  11$\sigma$, 8$\sigma$, $<$1$\sigma$.}
\end{figure}
\begin{figure}[t!]
\psfig{figure=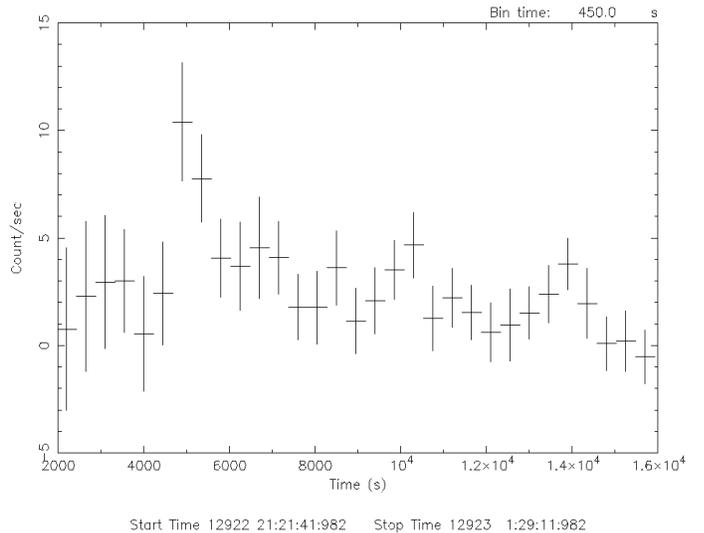,height=7cm,width=9cm}
\caption{ISGRI light curve (20--30~keV) of a newly discovered  outburst (No. 2 in Table 1) of SAX~J1818.6$-$1703.}
\end{figure}
SAX~J1818.6$-$1703 is a fast X-ray transient source which currently does not have any optical counterpart such as a galactic coronal source.
It was discovered and detected only once with the Wide Field Camera 2 onboard BeppoSAX on
11 March  1998 at RA=18$^{h}$ 18$^{m}$ 39$^{s}$ DEC=-17$^{\circ}$ 03$^{'}$ 06$^{''}$ (error radius 3$^{'}$)
(J. in't Zand et al. 1998). The source was seen to turn on at 19:12:00 UTC and peak quickly (20:38:24 UTC) 
at a level of 100 mCrab (2--9~keV) and 400 mCrab (9--25~keV). 
During  the last 40 minutes of the BeppoSAX observation, the source intensity decreased to an undetectable flux level.  
Although the WFCs  were operational up to April 2002,  monitoring the Galactic bulge once per week, no more outbursts 
of SAX~J1818.6$-$1703 detected by BeppoSAX have been reported in the literature.
Despite the fact that IBIS/ISGRI 
\begin{table*}[t!]
\begin{center}
\caption{Newly discovered outbursts of SAX~J1818.6$-$1703}
\begin{tabular}{clcccc}
\hline
\hline
No.  & Date  & Burst start time (UTC) & Burst stop time (UTC) & Energy Range & flux at the peak (20--30 keV)  \\
\hline
1   & 9 October 2003 & 13:06:48  & 14:08:57 & 20--60 keV & 178 mCrab \\  
2   & 10 October 2003 & 21:50:45 & 01:00:12 & 20--30 keV & 185 mCrab \\
\hline
\hline
\end{tabular}
\end{center}
\end{table*} 
performed numerous Core Program observations with the source in the field of view, to date 
SAX~J1818.6$-$1703 was not detected with INTEGRAL apart from
a weak detection reported by Revnivstev et al (2004) at a level of 5 mCrab in the energy range 18--60~keV, which was not reported as a transient event.
\begin{figure*}[t!]
\psfig{figure=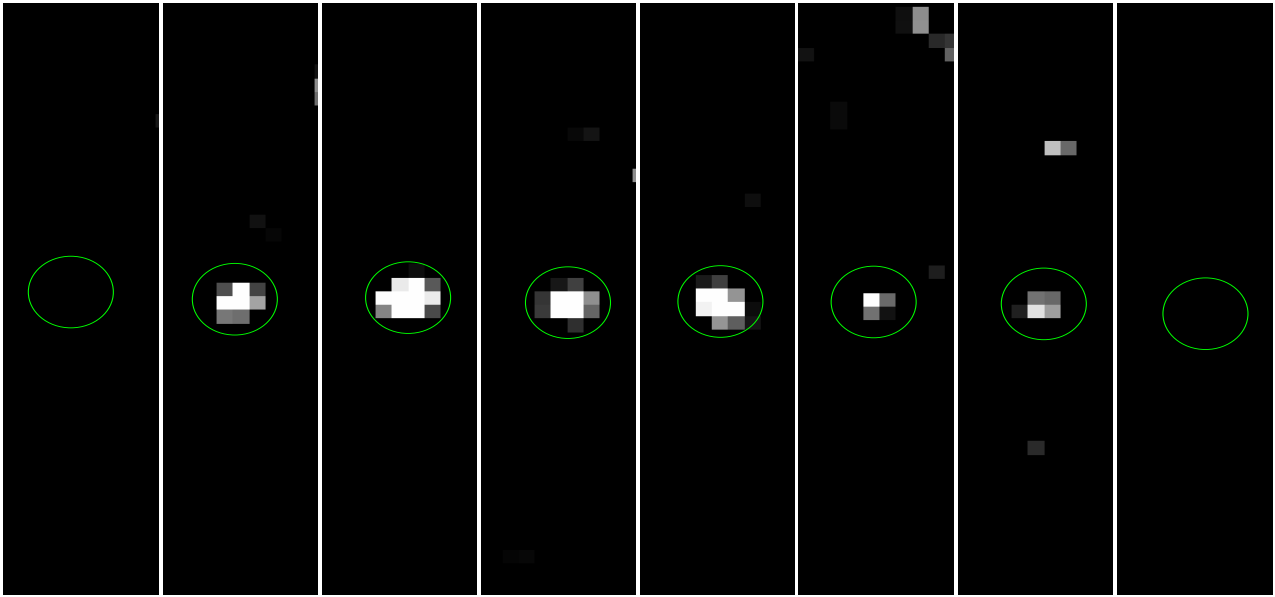,height=8cm,width=17cm}
\caption{ISGRI Science Window (ScW) image sequence (20--30 keV) of a newly discovered  outburst (No.2 in Table 1) of SAX~J1818.6$-$1703 (encircled).
The duration of each ScW is $\sim$ 2000 s.
The source was not detected in the first ScW on the left (significance less than 1$\sigma$), then it was 
detected during the next 6 ScWs with a significance, from left to right, equal to 5$\sigma$,  10.5$\sigma$,  7.5$\sigma$,  7$\sigma$,  5$\sigma$ and  
4.5$\sigma$, respectively. Finally in the last ScW the source was not detected (significance less than 2$\sigma$).}
\end{figure*}
\begin{figure*}[t!]
\psfig{figure=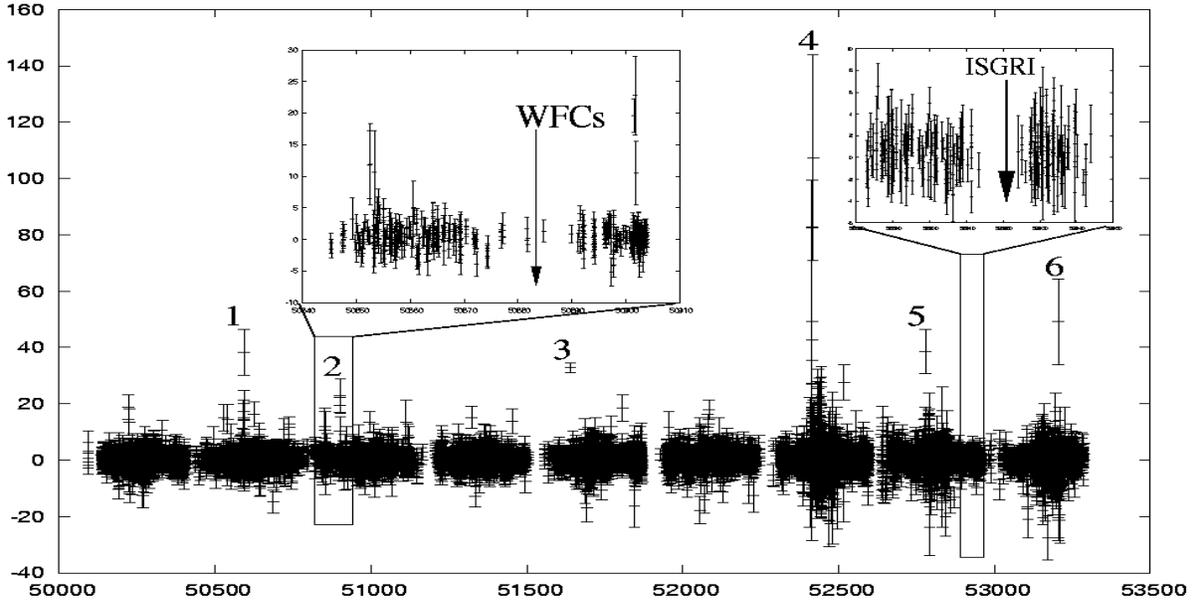,height=10cm,width=18cm}
\caption{RXTE ASM dwell by dwell light curve (2--12~keV) of SAX~J1818.6$-$1703 from 1996 to 2004. In the light curve 20 c s$^{-1}$ 
is equivalent to a flux of 250 mCrab. Time axis is in MJD.
The six outbursts detected by RXTE, with a flux greater than 250 mCrab, are indicated in the light curve by the numbers from 1 to 6. 
Zoomed views of the RXTE light curve during the epoch of the BeppoSAX WFCs and IBIS/ISGRI observations (marked by arrows) are shown in the insets, highlighting 
the lack of RXTE coverage during the outbursts.
The two ISGRI outbursts are indicated in the inset  on the right by means of only one arrow since they are
temporally very close each other.
ASM data can be retrieved on the public archive at
http://xte.mit.edu/XTE/asmlc/ASM.html}
\end{figure*}
\subsection{New results from analysis of IBIS/ISGRI observations}
We report on detection with IBIS/ISGRI of two outbursts of  SAX~J1818.6$-$1703, which are the first to be reported since 
its discovery with BeppoSAX WFCs in 1998.
The times of the 2 newly discovered  outbursts are listed in Table 1 together with  the energy range over which they have been detected and their fluxes at the peak 
in the 20--30~keV band.
We assumed the beginning of the first ScW during which the source was detected as being  the start time of the outburst and similarly the burst stop time to be the end of 
the last ScW during which the source was detected.

Outburst No. 1 (Table 1) was detected by ISGRI on 9 October 2003 only in 2 ScWs (which provides an upper limit on its duration of  $\sim$ 1 hour)
from 20 to 60~keV, with a  peak flux equal to $\simeq$ 180 mCrab 
(energy range 20--30~keV).
Figure 1 shows the sequence of consecutive ScWs (20--30~keV) during which it was observed.

Outburst No. 2 (Table 1) was detected only $\sim$1.5 days  later than No. 1. 
It was observed  in 6 ScWs (providing an upper limit on its duration of  $\sim$ 3.5 hours) only in the energy range 20--30~keV.
The source turned on at 10 October  2003  21:50:45 UTC and subsequently peaked after $\sim$ 35 minutes 
(10 October 2003 22:25:14 UTC) at a level of 185 mCrab. Subsequently the intensity decreased
to an undetectable flux level on 11 October 2003 01:00:12 UTC. 
Figure 2 shows its ISGRI light curve (20--30 keV), while Figure 3 shows  the sequence of consecutive ScWs (20--30 keV) during which it was observed.\\
We can note that these 2 newly discovered outbursts have a very similar peak flux value, both being $\simeq$ 180 mCrab in the  20--30~keV energy band. 
Moreover their duration (respectively $\sim$ 1 hour and $\sim$ 3 hours) is quite similar to that of the outburst detected by BeppoSAX WFCs when the source 
was discovered in 1998 ($\sim$ 2 hours).\\
An  investigation of the RXTE ASM (All Sky Monitor) data archive provided  a light curve (2--12~keV) of SAX~J1818.6$-$1703 from 1996 to 2004 (see Figure 4)
which shows at least 6 more outbursts having a flux greater than 250 mCrab in the energy range 2--12~keV (see numbers in Figure 4). 
A check of the RXTE light curve data points showed that there are none at the epoch of  
the  outbursts detected  by BeppoSAX WFCs and IBIS/ISGRI.
This can be clearly seen in the insets in Figure 4 which show a zoomed view of the RXTE light curve during the epoch of the BeppoSAX WFCs and IBIS/ISGRI observations,
marked by arrows.
It can be noted  that the outburst detected by BeppoSAX WFCs is very close to one of those detected by RXTE (outburst No. 2 in Figure 4), the former   
occurred 17.95 days earlier.\\
\section{IGR~J16479$-$4514}
\subsection{Archival X-ray observations of the source}
IGR~J16479$-$4514 was discovered with IBIS/ISGRI during observations of the Galactic Center region performed between August 8 
and August 9 2003 (Molkov et al. 2003), the measured fluxes were  12 mCrab and 8 mCrab in the energy bands 18--25~keV and 25--50~keV respectively. 
During observations performed on August 10, the source showed an increased outburst activity by a factor $\sim$2 in the energy bands 18--25~keV and 25--50~keV,
moreover it was also detected in the energy band 50--100~keV with a flux equal to  $\sim$ 17 mCrab (Molkov et al. 2003). 
To date this is the only outburst of IGR~J16479$-$4514 to be mentioned  in the literature.  This outburst was reported as an average detection during 
an observation of the Galactic center field  lasting about 3 days (Molkov et al. 2003), and  no analysis at the Science Window level was performed 
on this outburst, hence the fast transient nature of the source was not reported.      
Recently Lutovinov et al. (2004) published a broad band energy spectrum of IGR~J16479$-$4514 (1--100~keV) described by a simple power law
model modified by the cutoff at high energies and the photoabsorption at soft X-rays. 
The best fit parameters are a photon index of $\Gamma$=1.4$\pm$0.8, an observed absorption of N$_{H}$$\sim$1.2$\times$10$^{23}$ cm$^{-2}$ 
(exceeding the galactic value along the line of sight)
and a high energy cut-off having a value of E$_{c}$$\sim$32 keV. Based on these spectral characteristics, Lutovinov et al. (2004)
argue that IGR~J16479$-$4514 should be a neutron star binary system with high mass companion (HMXB).\\
The IGR~J16479$-$4514 error circle has a radius of  $\sim$ 3$^{'}$, no X-ray and radio sources have been found inside this radius 
using all available catalogues in the HEASARC database.
As for the infra-red band, Figure 5 shows the ISGRI error circle superimposed on the IRAS infra-red field (12 $\mu$m) as taken from the IRAS Sky Survey Atlas (ISSA) 
data archive. We note that on the edge of the ISGRI error circle there is a strong IRAS source (IRAS 16441$-$4506, located 2.9$^{'}$  from the ISGRI position)  
which is also identified with MSX5C G340.1271$-$00.0681 and a 2MASS infra-red source (see Table 2 for their fluxes). 
IRAS and MSX spectral bands (far infra-red) are essentially unaffected by dust or gas absorption, while on the contrary the 2MASS spectral band 
(near infra-red)  is strongly affected by absorption.
From Table 2, we can note that 2MASS flux measurements are much smaller than those of IRAS and MSX, 
which suggests that IRAS 16441$-$4506 could be a strongly absorbed source with a column density possibly  
exceding the value along the line of sight, which is $\sim$ 10$^{22}$ cm$^{-2}$ (Dickey \& Lockman 1990).
If IRAS 16441$-$450 is the infra-red counterpart of IGR~J16479$-$4514, this is in agreement with the high observed absorption of IGR~J16479$-$4514 (N$_{H}$$\sim$
1.2x10$^{23}$ cm$^{-2}$) reported by Lutovinov et al. (2004).
Moreover, IRAS 16441$-$4506  has been classified as an unusual source because its spectrum shows a strange and very strong peak emission near 11 $\mu$m (Volk et al. 1989).
In the IRAS image (Figure 5), an infra-red ring-like structure is clearly visible.
Apart from IRAS 16441$-$4506, the two other very strong IRAS sources 
(A and B in Figure 5) are classified as HII regions (SIMBAD). IGR~J16479$-$4514 seems to be located in an active  star formation region, unless this is a projection effect.
\begin{figure}[t!]
\psfig{figure=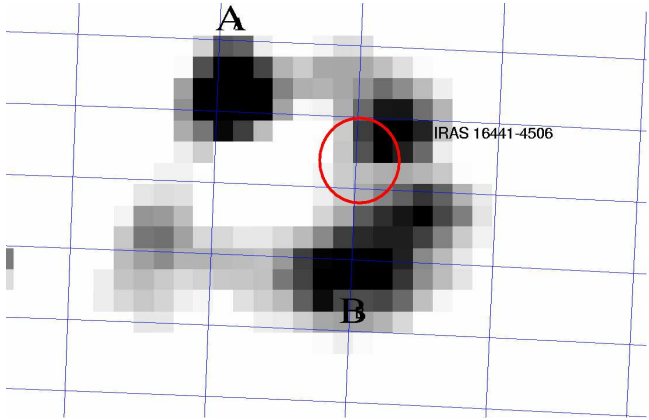,height=5.5cm,width=8.5cm}
\caption{IRAS infra-red picture (12 $\mu$m) superimposed on the ISGRI IGR~J16479$-$4514 error circle (3 arcmin radius).}
\end{figure}
\begin{table} [t]
\begin{center}
\caption{IRAS, MSX and 2MASS infra-red fluxes of the source IRAS 16441$-$4506}
\begin{tabular}{cc}
\hline
\hline
Wavelength & Flux \\   
($\mu$m) &   (Jy) \\  
\hline
IRAS 12 & 50.51 \\
IRAS 25 & 44.6  \\
IRAS 60 & 21.86 \\
IRAS 100 & 230.2 \\
MSX  12  & 55.46 \\
MSX 15 & 51.07 \\
MSX 21 & 46.82 \\ 
2MASS 1.25 & 0.005 \\
2MASS 1.65 & 1.14 \\
2MASS 2.17 & 5.37 \\ 
\hline
\hline
\end{tabular}
\end{center}
\end{table}
\subsection{New results from analysis of IBIS/ISGRI observations}
We report on 4 newly discovered outbursts of IGR~J16479$-$4514, derived from  analysis  of ISGRI data in revolutions number 47, 55, 63 and 101, respectively.
The times of the 4  outbursts are listed in Table 3 together with  the energy range over which they have been detected and their fluxes in the 20--30~keV band.
We assumed the beginning of the first ScW during which the source was detected as being  the start time of the outburst and similarly 
the burst stop time to be the end of the last ScW during which the source was detected.
We can note that their duration ranges from  $\sim$ 30 minutes to $\sim$ 3 hours, marking the very fast transient nature of IGR~J16479$-$4514, 
moreover they bracket in time the only outburst so far reported in the literature (8-10 August 2003).\\
Outburst No. 1 in Table 3 is particularly interesting, being by far the brightest one. Its ISGRI light curve (20--30~keV) is shown in Figure 6.
Initially the source flux is consistent with zero, then suddenly turns on at  13:36:37 UTC, flares up and quickly reaches the peak after
$\sim$ 5 minutes (13:41:39 UTC). Then
it drops to a very low flux level in $\sim$ 25 minutes and  continues to be characterized by a very low flux showing some 
secondary  peaks, as can be seen in Figure 6. At 16:58:30 UTC the source completely disappears.
The count rate at the peak is nearly  50 c s$^{-1}$ (energy band 20--30~keV) which  is equal to a flux of 850 mCrab 
or 3.8$\times$10$^{-9}$ erg cm$^{-2}$ s$^{-1}$.
Being located in the direction of the Norma region, an active star formation region,  we can assume a 
distance for IGR~J16479$-$4514 of 6 kpc, 
in this case the 20--30 keV luminosity is 1.6$\times$10$^{37}$ erg s$^{-1}$.
Figure 7 shows an expanded view of the outburst light curve binned  in 100 s periods, compared to the 250 s used 
in Figure 6. 
There is evidently a fast rise of the flare followed by a slower
decay during which there is an indication of a possible secondary peak.
We fitted an exponential function F=F(0)e$^{-t/\tau}$ to the observed flux count rate during the decay in the 20--30~keV energy band. 
This gives a $\chi^{2}_{\nu}$=0.8 for 15 d.o.f. and $\tau$=15$\pm$9 minutes, so it is reasonable to assert an exponential behavior for the decay of the outburst. 
\begin{table*} [t!]
\begin{center}
\caption{Newly discovered outbursts of IGR~J16479$-$4514}
\begin{tabular}{clcccc}
\hline
\hline
No.  & Date  & Burst start time (UTC) & Burst stop time (UTC) & Energy Range & flux (20--30 keV)  \\
\hline
1   & 5 March 2003 & 13:36:37  & 16:58:30 & 20--60 keV & 850 mCrab$\star$ \\  
2   & 28 March 2003 & 07:44:24 & 09:18:49 & 20--40 keV & 37 mCrab$^{\dagger}$ \\
3   & 21 April 2003 & 08:51:34 & 09:28:14 & 20--60 keV & 160 mCrab$^{\dagger}$\\
4   & 14 August 2003 & 00:19:29 & 02:07:44 & 20--60 keV & 44 mCrab$^{\dagger}$ \\
\hline
\hline
\end{tabular}
\end{center}
Note: $\dagger$ = average flux (20--30~keV) during the outburst \\
Note: $\star$ =  flux (20--30~keV) at the peak of the outburst \\
\end{table*} 

As for the remaining outbursts (No. 2, 3 and 4 in Table 3), they are shorter than No. 1 having been detected only in few ScWs (3, 1 and 3, respectively) 
during which their average
fluxes (20--30~keV) are respectively 37 mCrab, 160 mCrab and  44 mCrab.  
It can be noted that outburst No. 4 in Table 3 was detected a few days later ($\sim$ 5 days) than that
reported by  Molkov et al. (2003) when IGR~J16479$-$4514 was discovered.

We performed a Lomb-scargle periodogram analysis to search for periodicities using the ISGRI data of the four outbursts listed in Table 3, but no statistically 
significant periodicities have been found from 0.001 to 500 Hz.
\begin{figure}[t!]
\psfig{figure=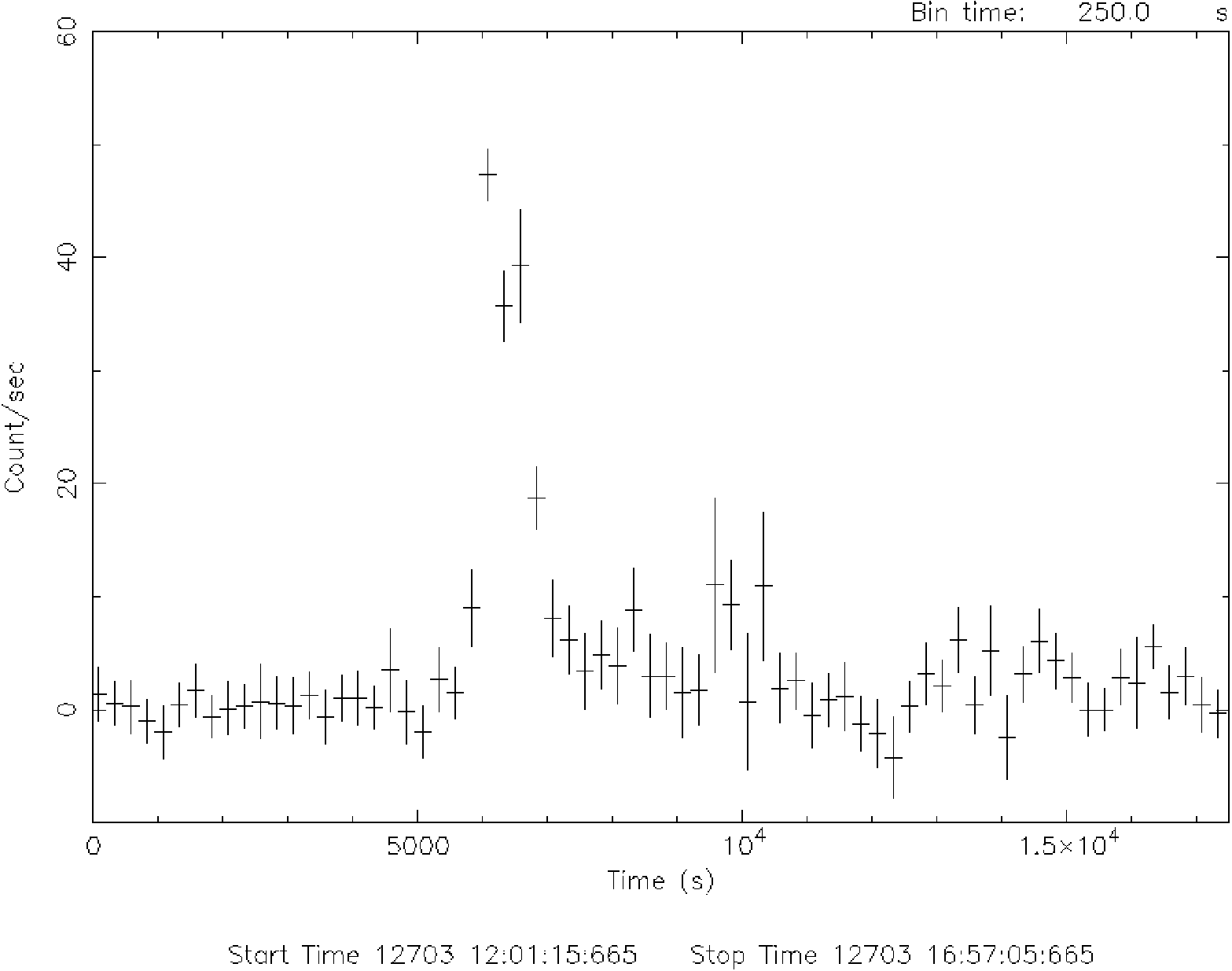,height=8cm,width=9cm}
\caption{ISGRI light curve (20--30~keV) of a newly discovered outburst (No.1 in Table 2) of IGR~J16479$-$4514.}
\end{figure}

Bearing in mind its spectral characteristics and its apparent location in a star formation region, IGR~J16479$-$4514 
could be a HMXB, as previously suggested by  Lutovinov et al. (2004). In the light of this interpretation, the 4 newly discovered outburts reported in Table 2
are possibly due to changes in the mass accretion rate onto the compact object (neutron star or black hole). 
The variations of the mass accretion rate can be related to an eccentric orbit, changes in the companion star
or instabilities of the accretion disk.  
However the durations of the four newly discovered outbursts (ranging from $\sim$ 30 minutes to  $\sim$ 3 hours) 
are very unusual for a HMXB given that it makes them 
significantly shorter than the type II outbursts of Be/NS binaries, which typically last for several weeks or even months.
\begin{figure}[t!]
\psfig{figure=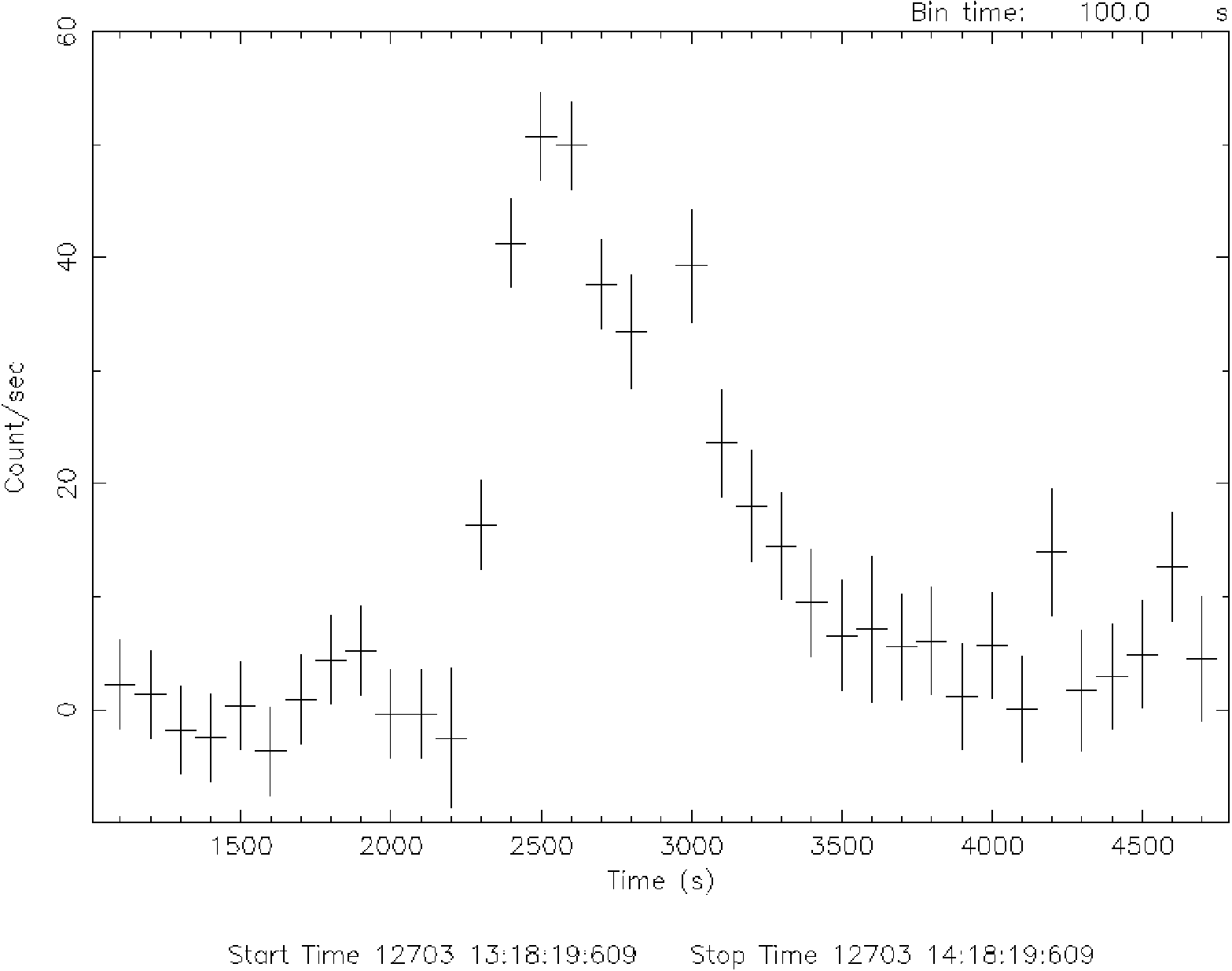,height=8cm,width=9cm}
\caption{Expanded view of the light curve (20--30~keV) of the outburst from IGR~J16479$-$4514 showed in Figure 6.}
\end{figure} 

We can note in Figure 7 that  at first glance the  time behavior of outburst No. 1 
(very rapid rise lasting  $\sim$ 5 minutes followed by a slower exponential decay lasting 
 $\sim$ 25 minutes) recalls a thermonuclear type I X-ray burst. About 40\% of the LMXBs in our Galaxy show type I X-ray bursts, which are 
explained as energy release by rapid nuclear fusion of material on the surface
of a neutron star. Thus a type I X-ray burst is thought to identify the source emitting it unambiguously as a LMXB containing a neutron star
(for reviews, see Lewin et al. 1993 and Strohmayer \& Bildsten 2003).
Normally  they are characterized by  rise times lasting from less than a second to $\sim$ 10 s, followed by a slower exponential decay lasting from  10 s to minutes.
During the decline they also show cooling of the characteristic temperature of the X-ray spectrum which is attributed to cooling of the neutron star surface.
The properties of a type I X-ray burst depend, according to theory, on
the mass and radius of the neutron star, on the rate with which  material is accreted onto the neutron star and on the composition of the accreted material.
Some longer type I X-ray bursts have been detected in many other sources.
In order of decreasing duration, some examples are: 4U1708-23, $\simeq$25 min (Hoffman et al. 1978, Lewin et al. 1984); 4U1724-307, $\geq$10 min (Swank et al. 1977);
3A1715-321, $\geq$4.5 min (Tawara et al. 1984a,b). 
It is worth noting that even longer type I X-ray bursts exist, the so-called superbursts, lasting for several hours (Cornelisse et al. 2000, 2002;
Wijnands 2001; Strohmayer \& Brown 2002; Kuulkers 2001; Kuulkers et al. 2002).   
Superbursts are more energetic and longer in duration than typical
type I X-ray bursts, but with similar spectral evolution. 
This suggests that they result from thermonuclear flashes occuring in fuel layers at much greater
depth than  typical type I X-ray bursts, thereby accounting for their longer duration.
The recurrence times of these events are not well constrained, so far only 7 events have been found, indicating that they are rare.\\
Typical type I X-ray outburst spectra are well described by a black body emission with temperature of $\sim$2--3 keV. The temperature
increases during the outburst rise (reflecting the heating of the neutron star surface) and decreases during the decay 
(due to subsequent cooling). We performed a spectral analysis  of the rise of outburst No. 1  as well as of its exponential decay, to find evidence of the cooling of 
the characteristic temperature of the X-ray spectrum, which would be a clear signature of a type I X-ray burst.
To this aim, first we created a user defined Good Time Interval (GTI) file to be used by OSA4.1 to
extract the spectrum of IGR~J16479$-$4514 during the rise and the exponential decay.
We rebinned the latest response matrix from the original 2048 channels to 64 channels,
the channels in the range 14--45 keV were grouped in bins of about 2 keV in size, while the ones in the range 45--100 keV were grouped in bins of about 5 keV size.
This was done to have enough statistics in the energy range 20--100 keV to perform the $\chi^{2}$ test in the spectral fit with XSPEC.\\
As for the 20--60 keV spectrum  extracted during the rise, the best fit model is a black body ($\chi^{2}_{\nu}$=1.6, d.o.f. 14)  with a temperature equal to
kT= 6.8$\pm$0.6 keV. Other spectral models, such as thermal bremsstrahlung or Comptonization models, provided instead very poor fits with a $\chi^{2}_{\nu}$ greater than 1.9.
The 20--60 keV spectrum extracted during the exponential decay can be fitted by black body model, giving a reasonable description ($\chi^{2}_{\nu}$=1.5, d.o.f. 14), with a 
temperature equal to kT= 6.4$\pm$0.3 keV. However a similarly reasonable fit was also achieved using a thermal bremsstrahlung or Comptonization model (CompST in XSPEC).
Clearly, the temperatures of the black body emission during the rise and the exponential decay are comparable within the errors of the 
measurements so the statistics are not good enough to find evidence of the cooling  of the characteristic temperature of the X-ray spectrum, 
which would be a clear signature of a type I X-ray burst. The black body spectrum is characterized by a  
temperature higher than typical type I X-ray bursts ($\sim$2--3 keV), although black body temperatures  in excess of 3 keV were observed 
for several sources (Lewin, Van Paradijs \& Taam 1993). Numerical model calculations of the radiative transfer in neutron star atmospheres show
that, although the shape of the spectrum is very nearly Planckian, in some cases the neutron stars are not very good black body emitters during the X-ray burst.   
The temperature fitted to the spectrum (i.e. the colour temperature T$_c$) can be higher than the effective temperature (T$_{eff}$) of the neutron star
atmosphere with values of the ratio  T$_c$/T$_{eff}$ as high as 1.7 (London et al. 1984, Foster et al. 1986, Ebisuzaki and Nomoto 1986).

On the one hand the energetic and the temporal behavior of the outburst (fast rise followed by a slower exponential decay) 
recalls a type I X-ray burst,  while conversely it was not possible to find evidence  of the cooling.
Hence a type I X-ray burst as  explanation for the outburst of IGR~J16479$-$4514 is inconclusive if only based on the temporal  behavior of the outburst.
If outburst  No. 1 is a type I X-ray burst, this would identify IGR~J16479$-$4514 as a LMXB containing 
a neutron star. However, this kind of explanation contrasts with the apparent source location in a star formation region  of the Norma spiral arm 
and so with a possible HMXB nature, 
as  was already proposed by  Lutovinov et al. (2004), 
unless the apparent location of IGR~J16479$-$514 in the Norma region  is a projection effect. To this aim, we inferred
an upper limit to the distance of the source. If it is indeed a type I X-ray burst, it should be Eddington limited with a peak flux
less than or equal to the Eddington limit, which for canonical neutron star values is   2$\times$10$^{38}$ erg s$^{-1}$. 
By setting the  Eddington limit equivalent to the bolometric observed 
flux we can calculate an upper limit to the distance, bearing  in mind that this kind of distance determination to IGR~J16479$-$4514 is only a rough estimate.
By doing so an upper limit  of $\sim$ 16 kpc was obtained. This value leaves the possibility that the location of 
IGR~J16479$-$4514 in a star formation region of the Norma spiral arm is only a projection effect. Unfortunately the value is  too high  to give any meaningful constraint 
to the distance and it might be due to the wrong assumption that the outburst is a Eddington limited type I X-ray burst.\\
\section{IGR~J17391$-$3021/XTE~J1739$-$302}
\subsection{Archival X-ray observations of the source}
The fast transient source IGR~J17391$-$3021 was discovered in outburst with the IBIS/ISGRI detector on 26 August  2003 at 18:49 UT (Sunyaev et al. 2003) with  a flux level of
70 mCrab in the energy range 18--50~keV. The maximum flux (150 mCrab) was detected on 27 August  2003 at 00:44 UT,  when  the source was also  detected
in the hard energy range 50--100 keV with a flux of 50 mCrab. The total length of the outburst was less than a day. The 20--100~keV spectrum 
of IGR~J17391$-$3021 at the maximum of the outburst was well fitted by an optically thin thermal bremsstrahlung model having  a temperature 
of T$\sim$22~keV (Lutovinov et al. 2005).
To date, this is the only outburst of IGR~J17391$-$3021 detected by IBIS/ISGRI and reported in the literature.\\
The IGR source position reported by Sunyaev et al. (2003), which has an error circle radius equal to 3$^{'}$, is located 
$\sim$1.5$^{'}$  from the Chandra location 
of the known transient XTE~J1739$-$302, which has a  position accuracy better than a few arcsec (Smith et al. 2004), 
so it cannot be excluded that IGR~J17391$-$3021 and XTE~J1739$-$302 are the same source.
The RXTE satellite pointed to XTE~J1739$-$302 $\sim$ 35 hours after the time quoted as the beginning 
of the previous cited outburst of IGR~J17391$-$3021 detected by ISGRI but 
no detection was reported (Smith et al. 2003).
\begin{table*} [t]
\begin{center}
\caption{Newly discovered outbursts of IGR~J17391$-$3021}
\begin{tabular}{clcccc}
\hline
\hline
No.  & Date  & Burst start time (UTC) & Burst stop time (UTC) & Energy Range & flux at the peak (20--30 keV)  \\
\hline
1   & 22 March 2003 & 12:12:44  & 14:00:44 & 20--60 keV & 254 mCrab \\  
2   & 9 March 2004 & 06:49:44 & 07:19:45 & 20--40 keV & 150 mCrab \\
3   & 9 March 2004 & 11:31:48 & ---   & 20--60 keV & 280 mCrab \\
4   & 10 March 2004 & 01:54:35 & 03:11:23 & 20--60 keV & 250 mCrab \\
\hline
\hline
\end{tabular}
\end{center}
\end{table*}
\begin{figure*}[t!]
\psfig{figure=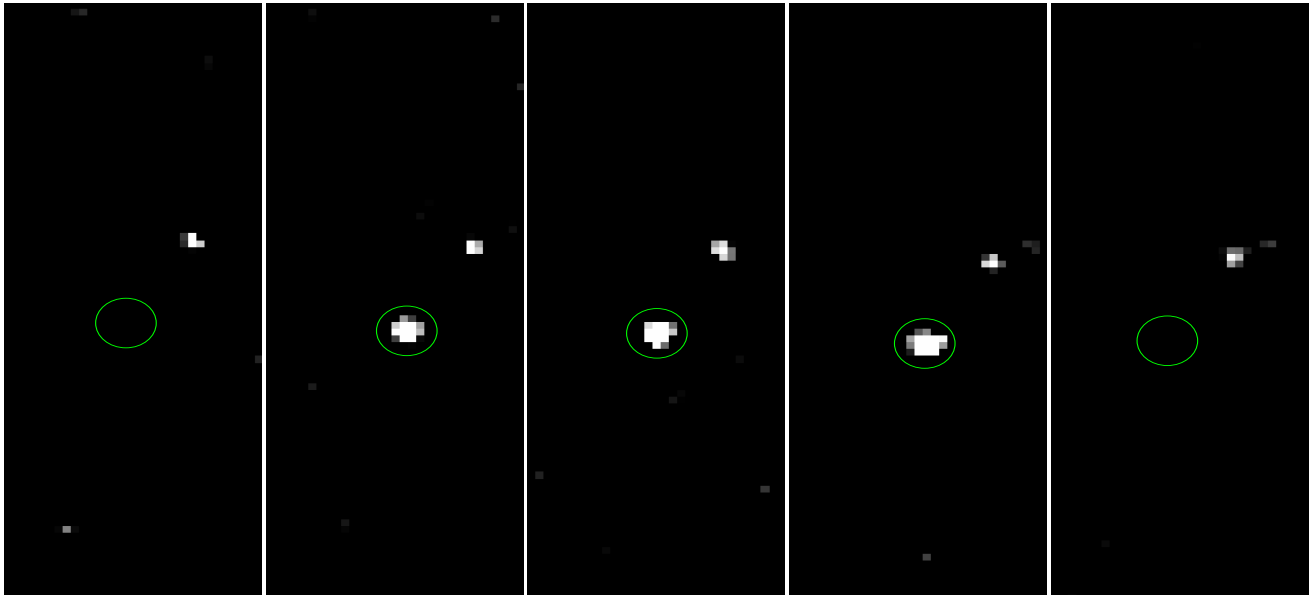,height=8cm,width=17.5cm}
\caption{ISGRI Science Window (ScW) image sequence (20--30~keV) of a newly discovered outburst (No. 1 in Table 4) of IGR~J17391$-$3021/XTE~J1739$-$302 (encircled). 
The duration of each ScW is $\sim$ 2000 s.
The source was not detected in the first ScW (significance less than 2$\sigma$), then
it was detected during the next 3 ScWs with a significance, from left to right, equal to 14$\sigma$,  16$\sigma$ and 23$\sigma$, respectively.
Finally in the last ScW the source was not detected (significance less than 2$\sigma$).
A weak persistent source (1E 1740.7-2942) is also visible in the field of view.}
\end{figure*}

XTE~J1739$-$302 (Smith et al. 1998) was the brightest source in the Galactic center region while active at the time
of the RXTE discovery (12 August  1997), with a flux of  3$\times$10$^{-9}$ erg cm$^{-2}$ s$^{-1}$ in the 2--25 keV band. 
The source was only observed by RXTE in that one day, it was 
not detectable 9 days earlier or 2, 8 and 16 days later. The RXTE spectrum during the bright state is well described by a thermal bremsstrahlung 
model with kT $\sim$ 12 keV, furthermore no statistically significant periodicities have been found from 0.01 to 1000 Hz.
Smith et al. (1998) tentatively identify XTE~J1739$-$302 as a Be/NS system,  
however  its outbursts (constrained by RXTE observations to be more than a  few hours but less than a  day) are much shorter 
than those typical of Be/NS binaries systems. Due to this unusual transient behavior, XTE~J1739$-$302
could define a new class of high mass fast X-ray transients (Smith et al. 2004). 

ASCA detected  XTE~J1739$-$302 on 11 March 1999  (Sakano et al. 2002), $\sim$ 1.5 years after the  RXTE discovery. 
The light curve (2--10 keV) initially showed no flux, then suddenly the source flared up reaching the peak after about 4 minutes, 
then dropped to zero flux level with the same timescale as the flare-rise. Then the source flared up again, this 
second flare being characterized by an almost identical profile except for the peak flux which was about half of that of the first flare. 
The 2--10 keV ASCA spectrum of XTE~J1739$-$302, accumulated during the flaring period, is a very hard and absorbed power law with an index corresponding to a
temperature over 100 keV when a  bremsstrahlung model is applied. Hence the ASCA spectrum is significantly different in shape and also the reported  
column density is lower than the one measured by RXTE in August 1997. 
ASCA timing analysis did not   find any significant pulsations, which is consistent with the RXTE results by Smith et al. (1998).

Recently, a Chandra observation of XTE~J1739$-$302  in quiescence provided  a very accurate position 
at RA=17$^{h}$ 39$^{m}$ 11.58$^{s}$ DEC=-30$^{\circ}$ 20$^{'}$ 37.6$^{''}$ with an position accuracy better than few arcsec (Smith et al. 2004). 
This permited the identification of its optical counterpart 
which is a bright highly reddened  
blue supergiant (O7.5-8Iaf) with no emission lines from an equatorial wind.
Despite its position near the galactic center region,  XTE~J1739$-$302 could  be a foreground object 
with a distance of about 1.8 kpc, as indicated by the bright optical companion. 
\begin{figure}[t!]
\psfig{figure=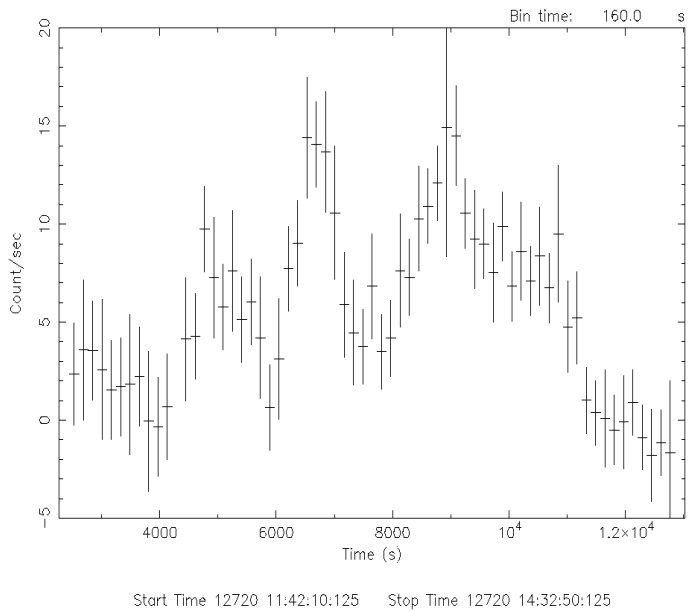,height=8cm,width=9cm}
\caption{ISGRI light curve (20--30 keV) of a newly discovered  outburst of IGR~J17391$-$3021 (No. 1 in Table 4)}
\end{figure}
\begin{figure}[t!]
\psfig{figure=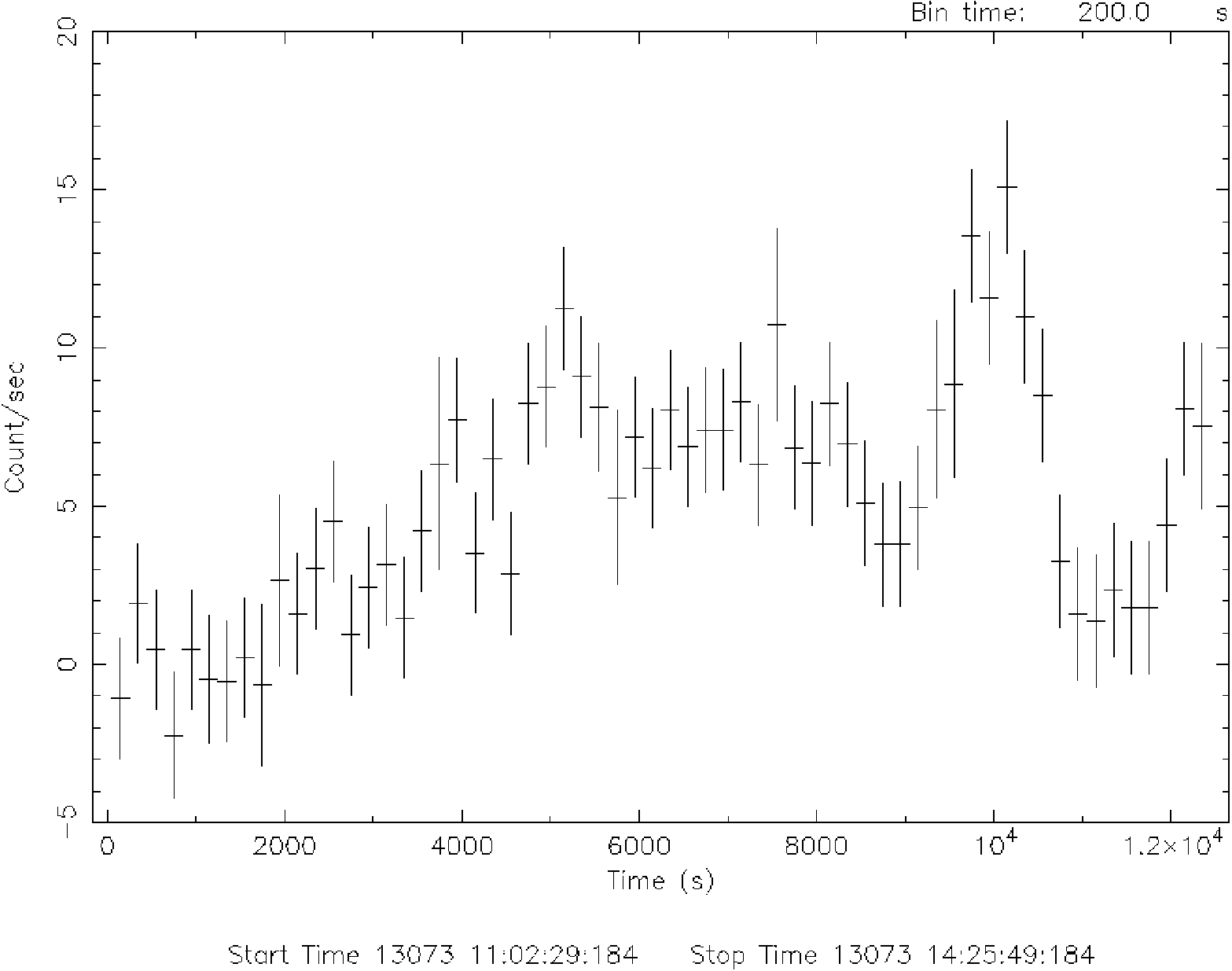,height=8cm,width=9cm}
\caption{ISGRI light curve (20--30 keV) of a newly discovered  outburst of IGR~J17391$-$3021 (No.3 in Table 4)}
\end{figure}
\begin{figure}[t!]
\psfig{figure=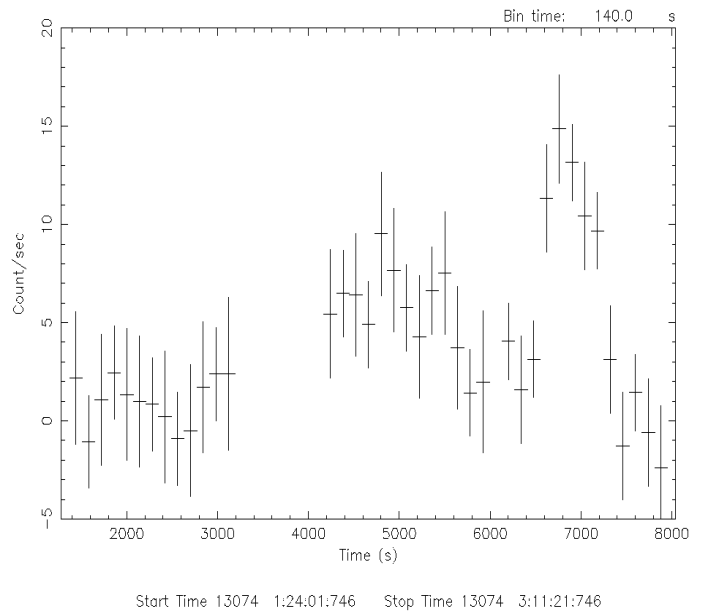,height=8cm,width=9cm}
\caption{ISGRI light curve (20--30 keV) of a newly discovered  outburst of IGR~J17391$-$3021 (No. 4 in Table 4)}
\end{figure}
\subsection{New results from analysis of IBIS/ISGRI observations}
We report on 4 newly discovered outbursts of IGR~J17391$-$3021 from analysis of ISGRI data in revolutions No. 53 and No. 171. 
Firstly, our ISGRI analysis enabled a more accurate
source position (RA=17$^{h}$ 39$^{m}$ 08.52$^{s}$ DEC=-30$^{\circ}$ 19$^{'}$ 55.8$^{''}$, error circle radius 1.3$^{'}$) than that reported by  
Sunyaev et al (2003) when the source was discovered.
Our improved position is now located  57$^{''}$  from the Chandra location of  XTE~J1739$-$302,
this significantly increases the probability that the 4 newly  discovered ISGRI outbursts and the one previously reported in the literature are from XTE~J1739$-$302.
The times of the 4 outbursts are listed in Table 4 together with the energy range over which they have been detected and their peak fluxes in the 20--30~keV band. 
We assumed the beginning of the first ScW during which the source was detected as being  the start time of the outburst and similarly the burst 
stop time to be the end of the last ScW during which the source was visible.
We  note that their durations range from $\sim$ 30 minutes to $\sim$ 2  hours, marking the very fast transient nature of IGR~J17391$-$3021.
The ISGRI light curves (20--30~keV) of the brightest outbursts (No. 1, 3 and 4  in Table 4) are shown in Figure 9, 10 and 11, respectively.

Outburst No. 1 lasted $\sim$ 2 hours and its ISGRI light curve (Figure 9) shows three noticeable peaks. 
The first two peaks have a very fast rise and decay as well as a similar duration, both being $\sim$ 20 minutes long.
The third peak lasts $\sim$ 1 hour showing a fast rise followed by a slower decay.
Furthermore, the last two flares have the same peak flux ($\sim$ 250 mCrab, energy band 20--30~keV) which is greater than that of the first flare  
($\sim$ 170 mCrab, energy band 20--30~keV).Figure 8 shows the sequence of 3 consecutive ISGRI ScWs during which this outburst has been detected.  
It is worth noting that this outburst was detected by JEM-X.
In the first and second ScW shown in Figure 8, during which the outburst was detected by ISGRI, the source is first outside and then on the 
edge of the JEM-X field of view so it was not detected.
On the contrary, during the last ScW, the source is inside the FOV of JEM-X and it was detected, although at  low  
significance (9$\sigma$, 9$\sigma$, 7$\sigma$ and 4$\sigma$ respectively in the energy ranges 3--6, 6--10, 10--15 and 15--35 keV). Unfortunately the statistics are not good  
enough to perform detailed spectral or timing analysis.       

Outburst No. 2 is the weakest and shortest one, being detected in only one ScW (providing an upper limit on its duration of  $\sim$ 30 minutes)
up to 40 keV. Its peak flux is 150 mCrab in the energy band 20--30~keV.

About five hours after outburst No. 2, the source turned on again (outburst No. 3 in Table 4). Its ISGRI light curve (Figure 10) shows 
at the beginning  a progressive rise lasting  $\sim$ 80 minutes during which the flux increases up to a value of 
$\sim$ 170 mCrab between 20--30 keV. Then it stops increasing and for $\sim$ 50 minutes it does not show 
any drastic increase or decrease. The source subsequently 
flares up quickly reaching a peak flux value of 280 mCrab and it  drops quickly with the same timescale as the rise-time ($\sim$ 15 minutes).
There is evidence of a possible second peak but unfortunately the light curve is truncated at 14:27:12 UTC 
because the following 2 ScWs are not available for data analysis.
In this case it is not possible to establish exactly when the outburst  terminates. However, we verified that  the source was not detected
in the next available ScW (starting at 15:24:01 UTC).\\
About 12 hours after outburst No. 3, IGR~J17391$-$3021 turned on again (outburst No. 4 in Table 4). In the light curve (see Figure 11) there is a gap due to
one unavailable ScW for data analysis, however the timing behavior of the source seems to suggest that it turned on 
just during the 
missing ScW, so it is reasonable to assume the outburst start time
at about 01:54:35 UTC. Once again, as seen in the previous outbursts, IGR~J17391$-$3021 shows a very 
quick flare with a peak flux equal to 
250 mCrab (20--30 keV), then it drops to an undetectable flux level with the same timescale as the time-rise ( $\sim$ 15 minutes). \\
A Lomb-scargle periodogram analysis was performed to search for periodicities using the ISGRI data  available
from the four outbursts reported in Table 4, 
but no statistically significant periodicity has been found.
\section{Conclusions}
We have presented an analysis of IBIS/ISGRI data on newly discovered outbursts of three fast X-ray transient sources.
The results confirm and
strengthen the very fast transient nature of these sources by verifying  that all the newly detected outbursts last less than a few hours. 
So far, FXTs did not show recurrence while all three fast transient sources  here reported
were detected in outburst by ISGRI more than once during the last 2 years, indicating for the first time a possible recurrent fast transient behaviour. 

In the case of the fast X-ray  transient source SAX~J1818.6$-$1703, we reported the detection with  ISGRI of two outbursts that occured 
only $\sim$1.5 days from each other.
Their durations  (respectively $\sim$1 hour and $\sim$3 hours) are quite similar to that of the outburst detected by BeppoSAX WFCs when the source 
was discovered in 1998 ($\sim$ 2 hours).

In the case of  IGR~J16479$-$4514, there might be reasons to consider it a HMXB, based on its spectral characteristics and its apparent location 
in a star formation region (Norma region).
However, the duration of the 4 newly discovered outbursts detected by ISGRI  and reported in this paper last less than 3 hours,
making them significantly shorter than the typical outbursts of HMXBs or Be/NS binaries. One of the four  outbursts is particularly interesting, its ISGRI light curve
(20--30 keV) shows a flare lasting $\sim$ 30 minutes and reaching a peak flux of 850 mCrab (luminosity of  1.3$\times$10$^{37}$ erg s$^{-1}$,
assuming the source is located in the Norma region). It is characterized by a rapid rise of $\sim$ 5 minutes followed by  a slower exponential decay of
$\sim$ 25 minutes. This 
timing behavior could be typical of 
a thermonuclear type I X-ray burst, identifing IGR~J16479$-$4514 as a LMXB containing a neutron star.
The spectrum of the outburst (20--60 keV) is fitted by a black body with a temperature
equal to $\sim$ 6  keV, which is  greater than that typical of type I X-ray bursts (2--3 keV), although black body temperatures  in excess of 3 keV were observed 
for several sources (Lewin, Van Paradijs \& Taam 1993). 
Moreover the statistics are not good enough  to find evidence of the cooling  of the characteristic temperature of the X-ray spectrum.
Hence this kind of explanation for the enhanced flux event of IGR~J16479$-$4514 is inconclusive if only based on time behavior of the outburst.

XTE~J1739$-$302 is an unusual transient source. First of all, our ISGRI analysis enabled a more accurate
IGR~J17391$-$302 position than that reported by  Sunyaev et al (2003) when the source was discovered.
Our improved position is located only 57 arcsec from the Chandra location of  XTE~J1739$-$302,
this significantly increases the probability that IGR~J17391$-$302 and XTE~J1739$-$302 are the same source.
The spectral characteristics of XTE~J1739$-$302 are typical of a neutron star binary, probably a HMXB given that its optical counterpart 
was identitifed with a blue supergiant. However, the duration of its outbursts are much shorter than those typical of HMXBs or Be/NS binaries systems, 
as confirmed by our reported ISGRI outburst detections with 
durations between 30 minutes and 3 hours. Although their ISGRI light curves 
show different timing behavior, they are all characterized by one or more quick flares having similar short duration ($\sim$ 15--20 minutes).  
An ASCA observation of XTE~J1739$-$302 in March 1999 provided a light curve (2--10 keV) showing the same kind of rapid flares.
XTE~J1739$-$302 could define a new class of high mass fast X-ray transients with blue supergiant secondaries (Negueruela et al 2005),
their fast outbursts could be perhaps due to  some kind of short ejection intrinsic to the secondaries. Another plausible 
explanation could be short viscous timescale in small accretion disks associated with wind accretion (Smith et al. 2004).
   
The unusual short and luminous outbursts of the three fast transient sources here reported make them particularly interesting.
This kind of sources are difficult to detect due to their very short outbursts, so 
it seems plausible that many such systems wait to be discovered in the Galaxy. Ongoing observations of IBIS/ISGRI may yield 
further detections of such sources.
\begin{acknowledgements}
V. Sguera is grateful to R. Cornelisse  and A. Tarana for useful discussions and suggestions and to J. L. Galache for providing the period analysis software. 
This research made use of data obtained from the HEASARC, SIMBAD and NED database. This research 
has been supported by  University of Southampton School of Physics and Astronomy.  AB, PU and AM thank
ASI financial support via  contract I/R/046/0. 
\end{acknowledgements}

\end{document}